\documentclass[useAMS,usenatbib]{mn2e}

\usepackage{graphicx}

\title[$\sigma$~Sco: Orbital Solution and Fundamental Parameters]
      {Orbital Solution \& Fundamental Parameters of $\sigma$ Scorpii}
      
\author[J. R. North et. al.]
       {J. R. North\thanks{E-mail: j.north@physics.usyd.edu.au},  
	J. Davis, P. G. Tuthill, W. J. Tango and J.G. Robertson \\
	School of Physics, University of Sydney, NSW 2006, Australia}

\begin{document}

\date{Accepted ; Received ; in original form }

\pagerange{\pageref{firstpage}--\pageref{lastpage}} \pubyear{2007}

\maketitle

\label{firstpage}

\begin{abstract}
The first orbital solution for the spectroscopic pair in the multiple
star system $\sigma$~Scorpii, determined from measurements with the
Sydney University Stellar Interferometer (SUSI), is presented.
The primary component is of $\beta$~Cephei variable type and has been
one of the most intensively studied examples of its class. The orbital
solution, when combined with radial velocity results found in the
literature, yields a distance of $174^{+23}_{-18}$\,pc, which is consistent
with, but more accurate than the {\em Hipparcos} value. For the primary
component we determine $18.4\pm5.4$\,M$_{\sun}$, $-4.12\pm0.34$\, mag and 
$12.7\pm1.8$\,R$_{\sun}$ for the mass, absolute visual magnitude and 
radius respectively. A B1 dwarf spectral type and luminosity class for the
secondary is proposed from the mass determination of $11.9\pm3.1$\,M$_{\sun}$
and the estimated system age of 10\,Myr.
\end{abstract}


%---------------------------------------------------------------------------%
%
%			      INTRODUCTION
%
%---------------------------------------------------------------------------%

\section{Introduction}

The class of variable stars with $\beta$~Cephei as the prototype consists 
of massive, nonsupergiant stars whose low-order pressure and gravity
mode pulsations result in light, radial and/or line profile variations
\citep{Stankov05}. In the most recent catalogue of $\beta$~Cephei type
variables, the 93 confirmed members have periods of strongest pulsation 
(some members display multi-mode pulsation behaviour) ranging from
1--8 hours and are of spectral type B0--B3 \citep{Stankov05}.

Approximately 14\% of the catalogued $\beta$~Cephei stars are located in 
multiple star systems and therefore, with sufficient observational data, 
component masses can be determined for comparison with those estimated from
theory. Indeed this has been achieved in the cases of $\alpha$~Virginis 
\citep{HEvans71}, $\beta$~Centauri (\citealt{Davis05}; \citealt{Auss06}) 
and $\lambda$~Scorpii \citep{Tango06} by combining the results of 
spectroscopic and interferometric analysis. Recently, analysis of the 
eclipsing binary HD~92024 has also yielded the mass determination of the 
$\beta$~Cephei primary \citep{Freyhammer05}. This last result is of 
particular interest as \citet{Freyhammer05} note that the HD~92024 
primary component spectrum closely resembles that of $\sigma$~Sco.

As a `classical' member \citep{Lesh78} of the $\beta$ Cephei variable 
type stars, $\sigma$~Sco (HR 6084, HD 147165) has been one of the most 
intensively studied examples of this class and has also been used as a 
photometric standard \citep{Vija93}.
\citet{Lutz77} classified the system as a binary with B2~IV + B9.5~V 
components separated by 20\arcsec \, on the sky and a visual magnitude 
difference of 5.31\,mag. Speckle interferometry, lunar occultations and
spectroscopy have since shown that the B2~IV component is in fact 
three stars -- a double-lined spectroscopic pair and a 2.2\, mag fainter 
B7 tertiary, 0.4\arcsec distant from the spectroscopic pair 
(\citealt{Beavers80}; \citealt{Evans86}; \citealt{Mathias91}).
The primary component has been identified as a $\beta$ Cephei type pulsator 
and has a classification of B1~III.
Therefore the system, as it stands in the literature, is quadruple:
a spectroscopic pair, a tertiary component, and the fourth star is the 
fainter component in the visual common proper motion pair ADS 10009
\citep{Pigulski92}.

The atmosphere of the $\beta$ Cephei component was analysed by
\citet{Linden88}, who gave an effective temperature of 
$26\,150\pm1\,070$\,K with a pulsation cycle variation of $4\,000\pm2\,000$\,K.
These authors also gave an excellent synopsis of early radial velocity and 
line profile studies of the spectroscopic pair. \citet{Mathias91} 
used a double-shock wave propagating in the stellar atmosphere to 
explain the observed {\em stillstand} in the radial velocity curves of the 
656.3\,nm H\,$\alpha$, 658.3\,nm C\,II and 455.3\,nm Si\,III lines
(a period of about an hour wherein the radial velocity remains relatively 
constant). The double-shock wave model was also successful in describing
similar characteristics of two other
$\beta$ Cephei stars: BW Vul and 12 Lac \citep{Mathias92}.
The study of \citet{Mathias91} was also the first to detect the 
spectral lines of the companion in the spectroscopic pair. However, neither 
the spectral type, luminosity class nor an estimate of the flux ratio  
(secondary/primary) was presented in \citet{Mathias91}. $\sigma$~Sco also 
exhibits a conspicuous 
{\em Van Hoof} effect -- a small phase lag of the radial velocity of the
hydrogen lines relative to that of all other lines (\citealt{Linden88};
\citealt{Mathias91}).

The most recent analyses of the spectroscopic pair by \citet{Mathias91}, 
\citet{Chapellier92} and \citet{Pigulski92} have produced orbital periods
of 33.012, 33.011 and 33.012 days respectively.
\citet{Pigulski92} has also concluded that the variations in the main 
pulsation period (increase in the first half of the 1900s, decrease after 
1960, and again an increase from about 1984) are due to a combination of 
evolutionary and light-time effects.

Hereafter, the following naming convention will be followed. The 
{\em primary} and {\em secondary} refer to the spectroscopic pair. The
{\em tertiary} refers to the star that is approximately 2.2\,mag fainter
and separated by 0.4\arcsec \, from the spectroscopic pair. The fourth and final
star will be referred to as the {\em distant companion}. The term 
$\sigma$~Sco will refer to the spectroscopic pair unless explicitly
stated otherwise.

In Section~\ref{fit} we provide the first complete orbital solution for 
$\sigma$~Sco determined by long-baseline optical interferometry. In 
combination with the only double-lined radial velocity measurements, the 
distance, spatial scale, mass and age of the components 
are quantified and compared to previous estimates in Section~\ref{sig_sys}. 
The details of our observations and a description of the parameter fitting 
procedure are described in Sections~\ref{obs} and \ref{fit} respectively.

%---------------------------------------------------------------------------%
%
%		      OBSERVATIONS AND DATA REDUCTION
%
%---------------------------------------------------------------------------%

\section{Observations and Data Reduction}
\label{obs}

Measurements of the squared visibility (i.e. the squared modulus of the 
normalised complex visibility) or $V^2$ were completed on a total of 31 
nights using the Sydney University Stellar Interferometer (SUSI,
\citealt{Davis99}). Interference fringes were recorded with the red 
beam-combining system using a filter with centre wavelength and full-width 
half-maximum of 700\,nm and 80\,nm respectively. This system was outlined by 
\citet{Tuthill04} and is to be described in greater detail by Davis et al. 
(in preparation). 

\begin{table}
    \caption{Adopted parameters of reference stars used during observations.
    The angular diameters (and associated error) were estimated from an 
    intrinsic colour interpolation (and spread in data) of measurements 
    made with the Narrabri Stellar Intensity Interferometer 
    \citep{HBrown74}.}
    \label{cal_table}
    \begin{tabular}{@{}cccccc@{}}
	\hline
    HR 	 & Name 	& Spectral & V    & UD Diameter       & Separation\\
	 &		&   Type   &      &  (mas)    	      & from $\sigma$ Sco\\
	\hline
    5928 & $\rho$ Sco   & B2IV/V    & 3.88  & $0.25 \pm 0.03$ &  6\fdg50 \\
    5993 & $\omega^1$ Sco &  B1Vp   & 3.96  & $0.26 \pm 0.04$ &  5\fdg93 \\
    6153 & $\omega$ Oph &  A7Vp     & 4.45  & $0.50 \pm 0.02$ &  4\fdg83 \\
    6165 & $\tau$ Sco   &  B0V      & 2.82  & $0.31 \pm 0.06$ &  4\fdg20 \\
	\hline
    \end{tabular}
\end{table}

The observations and data reduction followed the procedure outlined by
\citet{North07} using the adopted stellar parameters of the calibrator stars 
given in Table~\ref{cal_table}. A total of 262 estimations of $V^2$ were 
available for analysis (not including two observations that were removed 
due to possible tertiary contamination, Section~\ref{ter_effects}) as 
summarised in Table~\ref{obs_table}. 

\begin{table*}
  \centering
  \begin{minipage}{135mm}
    \centering
    \caption{Summary of observational data.  The night of the observation is 
             given in columns 1 and 2 as a calendar date and a mean MJD. 
	     Column 3 is the mean orbital phase calculated from the values in 
	     Table~\ref{tab:sig_Sco_final}. The baseline and the mean 
	     projected baseline (in units of metres) are given in columns 
	     4 and 5 respectively. Reference stars and the the number of 
	     squared visibility measures for a night are listed in the last two columns.} 
    \label{obs_table}
    \begin{tabular}{@{}lcccclr}
	\hline
	Date  	& MJD		&  Phase    & Nominal  & Projected & Reference & \# V$^2$\\
		&		&	    & Baseline & Baseline  &	Stars    &	   \\
	\hline
2005 May 20 & 53510.64 & 0.121 &  80 & 79.80 & $\tau$ Sco, $\rho$ Sco, $\omega$ Oph  & 4  \\
2005 May 24 & 53514.58 & 0.240 &  80 & 79.78 & $\tau$ Sco, $\rho$ Sco, $\omega$ Oph  & 10 \\
2005 May 25 & 53515.59 & 0.271 &  80 & 79.74 & $\tau$ Sco, $\rho$ Sco  & 4  \\

2005 July 19 & 53570.46 & 0.933 &  80 & 79.81 & $\tau$ Sco, $\rho$ Sco, $\omega$ Oph  & 10 \\
2005 July 20 & 53571.43 & 0.962 &  80 & 79.78 & $\tau$ Sco, $\rho$ Sco, $\omega^1$ Sco  & 11 \\

2005 August 08 & 53590.46 & 0.539 &  80 & 79.89 & $\tau$ Sco, $\omega^1$ Sco  & 11 \\
2005 August 11 & 53593.40 & 0.638 &  80 & 79.80 & $\tau$ Sco, $\omega^1$ Sco  & 8 \\
2005 August 13 & 53595.43 & 0.689 &  80 & 79.85 & $\tau$ Sco, $\omega^1$ Sco  & 10 \\
2005 August 16 & 53598.37 & 0.778 &  80 & 79.73 & $\tau$ Sco, $\omega^1$ Sco  & 2 \\

2006 June 15 & 53901.55 & 0.963 &  80 & 79.76 & $\tau$ Sco, $\rho$ Sco, $\omega^1$ Sco  & 3 \\
2006 June 16 & 53902.56 & 0.993 &  80 & 79.82 & $\tau$ Sco, $\rho$ Sco  & 9 \\
2006 June 17 & 53903.50 & 0.022 &  80 & 79.83 & $\tau$ Sco, $\rho$ Sco  & 12 \\
2006 June 18 & 53904.54 & 0.053 &  80 & 79.83 & $\tau$ Sco  & 15 \\
2006 June 24 & 53910.52 & 0.234 &  80 & 79.80 & $\tau$ Sco  & 11 \\
2006 June 25 & 53911.48 & 0.264 &  80 & 79.78 & $\tau$ Sco  & 8 \\
2006 June 26 & 53912.50 & 0.293 &  80 & 79.83 & $\tau$ Sco  & 3 \\
2006 June 27 & 53913.53 & 0.326 &  80 & 79.88 & $\tau$ Sco  & 8 \\
2006 June 28 & 53914.48 & 0.354 &  80 & 79.83 & $\tau$ Sco  & 12 \\

2006 July 18 & 53934.40 & 0.958 &  80 & 79.82 & $\tau$ Sco  & 2 \\
2006 July 19 & 53935.43 & 0.989 &  80 & 79.81 & $\tau$ Sco  & 11 \\
2006 July 25 & 53941.41 & 0.170 &  80 & 79.78 & $\tau$ Sco  & 8 \\

2006 August 01 & 53948.41 & 0.382 &  80 & 79.80 & $\tau$ Sco, $\rho$ Sco  & 12 \\
2006 August 03 & 53950.40 & 0.442 &  80 & 79.77 & $\tau$ Sco, $\rho$ Sco, $\omega^1$ Sco  & 8 \\
2006 August 04 & 53951.37 & 0.472 &  40 & 39.88 & $\tau$ Sco  & 4 \\
2006 August 07 & 53954.42 & 0.564 &  40 & 39.91 & $\tau$ Sco, $\rho$ Sco, $\omega^1$ Sco  & 13 \\
2006 August 08 & 53955.45 & 0.596 &  30 & 29.94 & $\tau$ Sco, $\rho$ Sco, $\omega^1$ Sco  & 12 \\
2006 August 09 & 53956.42 & 0.625 &  30 & 29.93 & $\tau$ Sco, $\rho$ Sco, $\omega^1$ Sco  & 12 \\
2006 August 10 & 53957.42 & 0.655 &  30 & 29.94 & $\tau$ Sco, $\rho$ Sco, $\omega^1$ Sco  & 12 \\
2006 August 22 & 53969.40 & 0.018 &  55 & 54.92 & $\tau$ Sco  & 6 \\
2006 August 26 & 53973.41 & 0.140 &  55 & 54.95 & $\tau$ Sco  & 6 \\
2006 August 27 & 53974.38 & 0.169 &  30 & 29.93 & $\tau$ Sco, $\rho$ Sco, $\omega^1$ Sco  & 5 \\
     \hline
    \end{tabular}
 \end{minipage}    

\end{table*}

%---------------------------------------------------------------------------%
%
%			    ORBITAL SOLUTION
%
%---------------------------------------------------------------------------%

\section{Orbital Solution}
\label{fit}

The theoretical response of a two aperture interferometer to the combined 
light of a binary star is given by \citep{HBrown70}
\begin{equation}
\label{binary_v2}
V^2 = \frac{V_1^2 + \beta^2V_2^2 + 2\beta |V_1||V_2| \cos(2\pi\bmath{b} \cdot \brho/\lambda)}
      { ( 1 + \beta ) ^2 },
\end{equation}
where $\beta < 1$ is the brightness ratio of the two stars in the observed 
bandpass and $V_1$, $V_2$ are the visibilities of the primary and 
secondary respectively.  
In the simplest case, stars can be modeled by a 
disc of uniform brightness with angular diameter $\theta$ (see 
Section~\ref{sig_radius} for the effects of limb-darkenening). The 
component visibilities in equation (\ref{binary_v2}) are then given by
\begin{equation}
\label{udisc_v}
V = \frac{2 J_1(\pi |\bmath{b}| \theta / \lambda)}
            {\pi |\bmath{b}| \theta / \lambda},
\end{equation}
where $J_1$ is a first order Bessel function.
The angular separation vector 
of the secondary with respect to the primary is given by $\brho$ (measured 
east from north), $\bmath{b}$ is the interferometer baseline vector 
projected onto the plane of the sky and $\lambda$ is the centre observing 
wavelength. The observed $V^2$ will vary throughout the night due to the 
orbital motion of the binary and Earth rotation of $\bmath{b}$. The 
Keplerian orbit of a binary star, i.e. $\brho$ as a function of time, can 
be parameterized with seven elements: the period $P$, eccentricity $e$, the 
longitude of periastron $\omega$, epoch of periastron $T_0$, semi-major 
axis $a$, the longitude of ascending node $\Omega$ and the inclination $i$. 
When using two-aperture optical interferometry, the phase of the complex 
visibility is lost and hence, $\Omega$ and $\omega$ have an ambiguity of 
180\degr. This is a direct result of the fact that component identities 
cannot be determined. Measurements of radial velocity can be used to 
remove the ambiguity of $\omega$ but that of $\Omega$ remains.

%---------------------------------------------------------------------------%
\subsection{Effects of the Tertiary and Distant Companion} 
\label{ter_effects}

The presence of a third or fourth component can greatly complicate 
analysis of interferometric data. The distant component is too far away 
and too faint to affect the interferometric data and can be neglected.
Therefore, an investigation into the effects of the tertiary star was 
conducted to validate the data and aid the final analysis.

\begin{figure}
     \includegraphics[width=\linewidth]{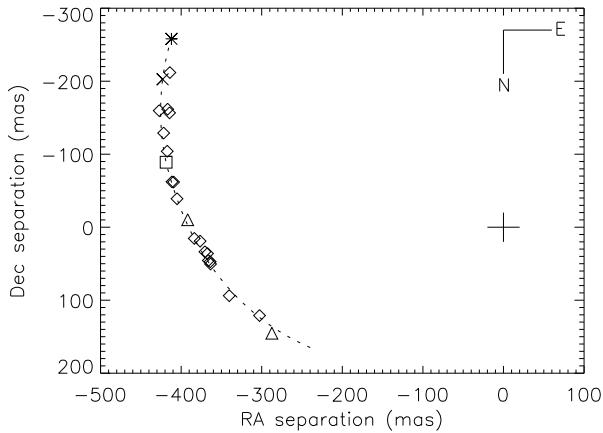}     
	\caption{Literature vector separations of the tertiary measured
	         with respect to the spectroscopic pair. Lunar occultations
		 are given as triangles, speckle interferometry as diamonds, 
		 {\em Hipparcos} results as a square and the coronographic 
		 measurement as a cross. The plus symbol represents the 
		 location of the spectroscopic pair. The dotted line is a 
		 simple smooth trajectory fit to the data and the 
		 asterisk is the estimated mean position of the tertiary 
		 during SUSI observations.}
        \label{fig:sig_Sco_ter}	
\end{figure}

Firstly, the mean location of the tertiary, relative to the spectroscopic
pair, was estimated by fitting a simple smoothed curve trajectory to the 
published vector separations. The literature vector separations are given 
in Table~\ref{tab:sig_Sco_ter} and are shown with the fitted curve in 
Fig.~\ref{fig:sig_Sco_ter}. The estimated vector separation for the
mean observing date of the SUSI observations (B2006.357) is $\rho=486$\,mas
and $\theta=238^{\circ}$ where $\rho$ and $\theta$ are the magnitude
and angle of the separation vector.
The magnitude difference of the tertiary with respect to the combined
spectroscopic pair irradiance is approximately $2.23\pm 0.18$\,mag (mean 
and standard deviation of the values from \citealt{Hartkopf06} adjusted 
to our passband). 
Given the estimated separation (which is within SUSI's 
field-of-view), the irradiance of the tertiary star will be detected by 
SUSI and must be included in the analysis. 
\begin{table}
    \centering
    \caption{Literature vector separations of the tertiary measured
	     with respect to the spectroscopic pair. Methods are
	     O: Occultation; S: Speckle; H: {\em Hipparcos}; 
	     C: Coronograph }
    \label{tab:sig_Sco_ter}

    \begin{tabular}{lccl}
	\hline
Besselian 	& $\rho$ 	& $\theta$	& Method:  Reference \\
Year	 	&  (mas)  	& 	(deg)	&	 \\
	\hline
1972.556	&	322	&	296.8	& O: \citet{Evans86} \\
1976.471	& 	326	&	291.8	& S: \citet{Morgan78}\\
1977.4868	& 	353	&	285.4	& S: \citet{McAlister79}\\
1980.4792	& 	367	&	277.4	& S: \citet{McAlister83}\\
1980.4819	& 	367	&	277.9	& S: \citet{McAlister83}\\
1981.4567	& 	372	&	275.2	& S: \citet{McAlister84}\\
1981.4704	&	369	&	277.1	& S: \citet{McAlister84}\\
1981.4730	&	369	&	275.6	& S: \citet{McAlister84}\\
1983.4254	&	377	&	272.6	& S: \citet{McAlister87}\\
1984.3783	&	384	&	272.3	& S: \citet{McAlister87}\\
1986.243	&	392	&	268.5	& O: \citet{Evans86} \\
1987.2726	&	407	&	264.5	& S: \citet{McAlister89}\\
1989.2275	&	416	&	261.4	& S: \citet{McAlister90}\\
1989.3038	&	414	&	261.4	& S: \citet{McAlister90}\\
1991.3140	&	428	&	258	& H: \citet{ESA97}\\
1993.3424	&	430	&	256	& S: \citet{Miu95}\\
1994.3509	&	441	&	253	& S: \citet{Mason96}\\
1997.2252	&	443	&	249.3	& S: \citet{Horch99}\\
1997.5174	&	465	&	242.9	& S: \citet{Horch99}\\
1997.5174	&	456	&	249.5	& S: \citet{Horch99}\\
1997.6157	&	447	&	248.8	& S: \citet{Horch99}\\
2000.3996	&	469	&	244.4	& C: \citet{Shatsky02}\\
	\hline
    \end{tabular}
\end{table}

The offset of the tertiary fringe packet in delay space, $D$, was calculated 
using 
\begin{equation}
\label{eq:sig_sco_offset}
    D = | {\bf b} | \rho \cos\left(\eta - \theta \right),
\end{equation}
assuming that the spectroscopic pair's fringe packet is located at 
the phase centre of SUSI. $|{\bf b}|$ and $\eta$ are the projected baseline 
length and position angle respectively. It was found that only two 
observations had an offset 
within the centre half of the 140\,$\mu$m observing scan. For 
approximately 72\% of the observations, the offset was sufficiently large to
place the tertiary fringe packet entirely outside the observing scan. 
However, the SUSI data reduction pipeline windows the recorded scan about the 
detected peak fringe location \citep{Ireland05}. Therefore, the tertiary 
fringe packet will not be within the `data window' and should not affect the 
calculation of $V^2$ if both
\begin{enumerate}
\item the offset is sufficiently large, and
\item the fringe packet of the spectroscopic pair has a greater amplitude 
      than that of the tertiary (i.e.\ avoid the system `locking' onto the 
      tertiary).
\end{enumerate}
The first condition can easily be satisfied by the removal of all 
observations with an offset less than 30\,$\mu$m -- the outer quarter of 
a scan (beyond the edge of the window function).
Hence, the two data points that were found within the centre half of the 
fringe scan were rejected from the analysis. The second condition may not be 
satisfied when the primary and secondary fringe packets destructively 
interfere; i.e.\ at a minimum in the modulation of $V^2$ there is the 
possibility of mistakenly measuring the interference pattern of the tertiary.
Assuming the primary and secondary 
fringe packet destructive interference is absolute and the tertiary is 
completely unresolved, then the expected $V^2 \la 0.013$ will be due to the 
tertiary alone. As this is below the $V^2$ detection limit of SUSI, we can
neglect such cases.
Hence all remaining observations of $\sigma$~Sco now satisfy both conditions.
The SUSI data pipeline will nevertheless be affected by the incoherent 
flux from the tertiary component. The required adjustment to the fitted 
squared-visibility model is given in the next section.

%---------------------------------------------------------------------------%

\subsection{Fitting Procedure and Uncertainty Estimation}
\label{fit_err}

The validity of equation (\ref{binary_v2}) is (strictly) only for 
observations of binary stars made with very narrow bandwidths.  
For real detection systems {\em wide bandwidth effects} can reduce the 
observed $V^2$ and for the scanning detection system of SUSI, the 
equivalent to equation (\ref{binary_v2}) giving $V^2$ for a binary star
for the case of a wide spectral bandwidth is \citep{North07}:
\begin{equation}
\label{wide_v2}
V^2 = \frac{V_1^2 + \beta^2 V_2^2 + 
	    2\beta r(\psi)|V_1||V_2| \cos(\psi)}
    	    { ( 1 + \beta ) ^2 },
\end{equation}
where
\begin{equation} 
r(\psi) = \exp \left[\frac{-\Delta\lambda^2}{\lambda_0^2}\frac{\psi^2}{32\ln2}\right].
\end{equation}
The spectral response is approximated as a Gaussian of centre 
wavelength $\lambda_0$ with full-width half-maximum $\Delta\lambda$.
The term $r(\psi)$ corresponds to the autocorrelation of the Gaussian
envelope of the interference pattern and  
$\psi = 2\pi\bmath{b} \cdot \brho/\lambda_0$ is defined for convenience. 

The measures of $V^2$ are contaminated by the irradiance of the tertiary
star (Section~\ref{ter_effects}) such that equation (\ref{wide_v2}) is no 
longer applicable. The brightness ratio $I_3$ of the 
tertiary/(primary + secondary) is approximately $I_3 \simeq 0.128\pm0.023$
(Section~\ref{ter_effects}). Adjusting equation (\ref{wide_v2}) for the 
contamination of the tertiary we obtain
\begin{equation}
\label{eq:sig_Sco_V2}
|V|^2 = \frac{V_1^2 + \beta^2 V_2^2 + 
	2\beta |V_1| |V_2| r(\psi) \cos(\psi) }
	{[(1+I_3)(1+\beta)]^2}.
\end{equation}
The term, $I_3$, reduces the observed $V^2$ and hence the effect the 
tertiary component has on the calculated $V^2$ of the spectroscopic pair is 
considered an extra incoherent source. The addition of this term
will mainly affect the fitted component angular diameters and brightness
ratio, leaving the orbital parameters relatively unaffected.
As $\Delta\lambda$ and $I_3$ approach zero, i.e. narrow bandwidth
observations of a simple binary star, then equation (\ref{eq:sig_Sco_V2}) 
reduces to equation (\ref{binary_v2}). 

Initial values of the inclination and position angle of the ascending node
were found by a coarse grid search of parameter space. The remaining orbital
parameters were limited to within three standard deviations of the
values given by \citet{Mathias91}. The inital angular diameter of the
primary star was estimated from the {\em Hipparcos} distance and spectral
type characteristics found in the literature (\citealt{Cox00}).
By physical arguments and inspection of the measured $V^2$ values, the
secondary's angular diameter and brightness ratio were limited to the
ranges 0.2--0.6\,mas and 0.2--0.8 respectively.

The final estimation of parameters was completed using $\chi^2$ minimization
as implemented by the Levenberg-Marquardt method to fit equation 
(\ref{eq:sig_Sco_V2}) to the observed values of $V^2$. 
When finding the minimum of the $\chi^2$ manifold, the inverse of the 
covariance matrix is calculated by the non-linear fitting
program.
The formal uncertainties of the fitted parameters are derived from
the diagonal elements of this covariance matrix.
As the visibility measurement errors may not strictly conform to a normal 
distribution and equation (\ref{eq:sig_Sco_V2}) is non-linear, the formal 
uncertainties may be underestimates. Following the approach of
\citet{North07}, three uncertainty estimation methods were adopted to
confirm the accuracy of the values derived from the covariance matrix.
These methods: Monte Carlo, bootstrap and Markov chain Monte Carlo 
(MCMC) simulations are described by \citet{North07} and references therein.

%---------------------------------------------------------------------------%
\subsection{Results}
\label{res}
\begin{table*}
    \begin{minipage}{145mm}
    \centering
    \caption{Orbital and physical parameters of $\sigma$~Sco as found with 
	     SUSI and from the literature. The periods from the literature
	     have been converted to MJD and/or adjusted to the same epoch.
	     Note that the difference between the Heliocentric Julian Day 
	     (HJD) and a Julian Day is less than ten minutes which is 
	     much smaller than the uncertainties.}
    \label{tab:sig_Sco_final}
        \centering
    \begin{tabular}{lcccc}
	\hline
Parameter & Unit   & 	     SUSI	&   \citet{Mathias91} & \citet{Pigulski92} \\
	\hline
$P$ 	  & days   & $33.010 \pm 0.002$ &   $33.012 \pm 0.002$	&   	$33.012$ 	\\
$e$ 	  &	   &$0.3220 \pm 0.0012$	&      $0.44 \pm 0.11$ 	&      $0.40 \pm 0.04$  \\
$T_0$     & MJD	   &$34889.0 \pm 1.0$	& $34888.9 \pm 0.7^b$ 	&  $34888.0 \pm 0.4^c$	\\
$\omega$  & deg	   &	$283 \pm 5 ^d $	&  $299.1 \pm 10.0$	&       $287 \pm 6$ 	\\
$a''$  	  &  mas   &   $3.62 \pm 0.06$  &           -		&      -  		\\
$\Omega$  & deg	   &	$104 \pm 5 $	&        -		&       -		\\
$i$       & deg	   &  $158.2 \pm 2.3$	&       -	  	&      - 		\\
$\theta_1$& mas    & $0.67 \pm 0.03$	&	-	 	&       -		\\ 
$\theta_2$& mas    & $0.34 \pm 0.04^a$	&	-	 	&      -   		\\
$\beta$	  &	   & $0.48 \pm 0.02$	&	-		&     -   		\\
	\hline
    \end{tabular}\newline \centering
    $^a$ Adopted parameters;
    $^b$ HJD-2400000.5;
    $^c$ 498 $\times$ 33.012d added;
    $^d$ Ambiguity of 180\degr.
    \end{minipage}
\end{table*}

\begin{figure*}
    \begin{minipage}{145mm}
    \includegraphics{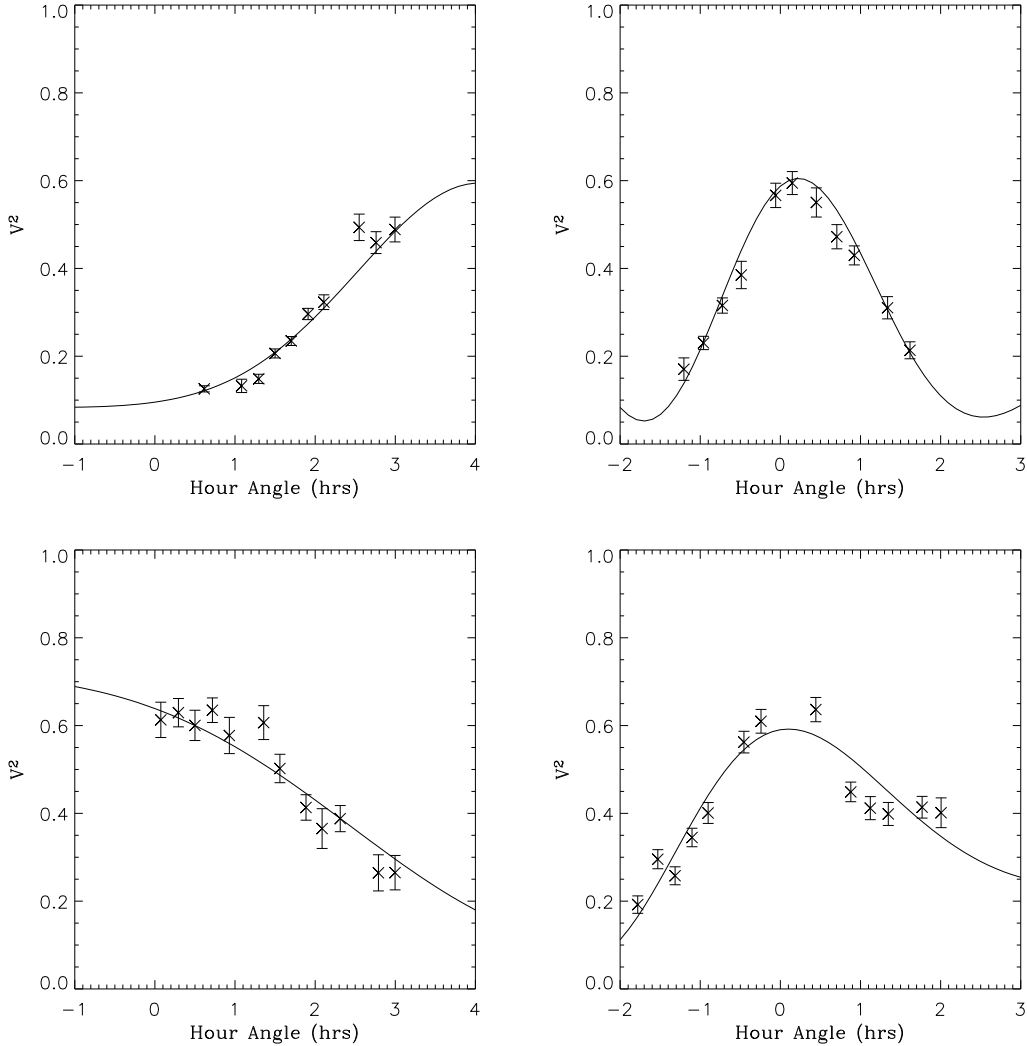}
    \caption{Data from the nights of 2005 Aug 08 (top-left), 2006 Aug 08 
	     (bottom-left), 2006 Jun 24 (top-right) and 2006 Jun 28 
	     (bottom-right) where each data point represents a measure of 
	     $V^2$ with the associated formal error. The values of 
	     Table~\ref{tab:sig_Sco_final} have been been used to show 
	     the fitted model as a solid line.}
    \label{fig:sig_Sco_example} 
    \end{minipage}
\end{figure*}

From preliminary analysis, a classification of B1~V was adopted for the 
secondary (see Section~\ref{sig_sec}) and its angular diameter was assumed 
to be $0.34\pm0.04$\,mas (i.e.\ a radius of 6.4\,R$_{\sun}$ -- with a 
12 per cent uncertainty -- at the distance obtained from the dynamical 
parallax). The three parameters $\theta_1$, $\theta_2$ and $\beta$ are 
coupled and hence a change in one will affect the other two without 
significantly changing the orbital parameters. 

\begin{figure}
 \includegraphics[width=\linewidth]{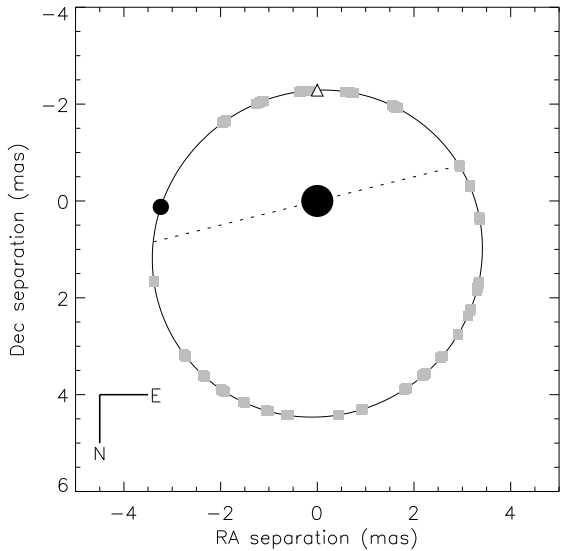}
 \caption{The relative orbit (solid line) of the secondary about the 
 $\beta$~Cephei star projected on the plane of the sky. The dotted 
 line is the line-of-nodes, the open triangle signifies periastron and the 
 grey squares are parts of the orbit that were observed with SUSI. The 
 stellar components (filled circles) are drawn to scale using the values of 
 Table~\ref{tab:sig_Sco_final}. The secondary is shown at an orbital phase 
 of 0.85. }
 \label{fig:sig_Sco_orbit} 
\end{figure}

The best-fitting values of the model parameters are given in 
Table~\ref{tab:sig_Sco_final} and four nights data are shown in 
Fig.~\ref{fig:sig_Sco_example} with the predicted $V^2$ model overlaid 
as the solid curve. The projected orbit on the plane of the sky is shown in 
Fig~\ref{fig:sig_Sco_orbit}.

The reduced $\chi^2$ of the fit was 3.58, implying that the measurement 
errors are underestimated by a factor of 1.89. Two possible effects were
investigated. Firstly, the seeing conditions during some observations were 
poor and then only the brightest calibrator could be used, resulting in 
data of lower quality than on other nights. While a non-linear seeing 
correction \citep{Ireland06} was applied to all data as part of the data 
reduction, there could be some residual atmospheric effects. Secondly, the
primary star, being of $\beta$ Cephei pulsator type, varies in diameter,
temperature and apparent brightness. These properties will also affect the 
results of fitting equation (\ref{eq:sig_Sco_V2}) to the data and contribute 
to the somewhat large value of reduced $\chi^2$ and non-Gaussian
parameter uncertainty distributions (discussed further below).
However the quality and resolution of the SUSI data combined with the 
uncertainty in the literature values does not justify a more 
complicated model that includes the intrinsic variability. 
Furthermore, the number and time base of $V^2$ measures is sufficient to 
average out any effects from the intrinsic variability of the primary. 
Hence, the angular diameter of the primary, along with the component 
brightness ratio, are treated as mean values by 
equation (\ref{eq:sig_Sco_V2}).

The three different uncertainty estimation techniques (described in
Section~\ref{fit_err}) produced distributions of each free model parameter
similar in appearance and approximately centred on the best-fitting values.
The Monte Carlo and 
bootstrap methods were set to each generate $10^3$ synthetic data sets 
while the MCMC simulations completed $10^7$ iterations. The shape of the 
likelihood functions of the semi-major axis, eccentricity, inclination, 
uniform disc angular diameters and the brightness ratio were Gaussian in 
appearance. The Probability Density Functions produced by the MCMC 
simulation for the remaining (free) model parameters were slightly 
non-Gaussian, most likely due to the intrinsic pulsations of the primary. 
Even though some model 
parameters produced (weakly) non-Gaussian distributions, the uncertainty 
values quoted in Table~\ref{tab:sig_Sco_final} are the standard deviations. 
As the MCMC simulation includes the uncertainties in the tertiary incoherent 
flux and angular diameter of the secondary, the values it produced form the 
basis of the final parameter uncertainties given in 
Table~\ref{tab:sig_Sco_final} because we believe they are the most realistic 
estimates of parameter uncertainties for our data set.

The orbital parameters found by the analyses of \citet{Mathias91} and
\citet{Pigulski92}, together with the final values determined from the 
SUSI data, are given in Table~\ref{tab:sig_Sco_final}. The values for the
period and time of periastron passage are all in excellent agreement.
There is some disagreement among the two remaining orbital parameters
when considering the quoted uncertainties. However, all values are 
consistent at the two standard deviation level.
The analysis of \citet{Mathias91} is considered to be the most
up-to-date due to the detection of the secondary spectral lines and the 
inclusion of the most recent (published) data.
Therefore, comparing only the SUSI and the \citet{Mathias91} results
all parameters are consistent at the 1.1 standard deviation level.

%---------------------------------------------------------------------------%
%
%		     SYSTEM AND PHYSICAL PARAMETERS
%
%---------------------------------------------------------------------------%

\section{System and Physical Parameters}
\label{sig_sys}

\citet{Mathias91} contains the only published semi-amplitudes of both
the primary and secondary components. The small discrepancy between the SUSI 
and \citet{Mathias91} eccentricity and longitude of periastron passage may 
affect the estimation of those physical parameters obtained by the 
combination of interferometric and spectroscopic results. We note, however, 
that the inclination is close to 180\degr and consequently, the uncertainty 
in $\sin i$ will dominate the error budget (see for example 
Section~\ref{sig_dis}).

%---------------------------------------------------------------------------%
\subsection{Distance}
\label{sig_dis}
The calculation of the dynamical parallax, 
\begin{equation}
\label{dyn_par}
\pi_d = \frac{a\arcsec}{(a_1 + a_2)},
\end{equation}
requires the semi-major axis of the relative orbit in both linear units 
(i.e.\ AU) and angular units. Using the interferometric orbital values and 
the component semi-amplitudes $K_1 = 31.9\pm1.3$\,kms$^{-1}$ \& 
$K_2 = 49.3\pm7.3$\,kms$^{-1}$\citep{Mathias91}, the semi-major axes of the
component orbits were calculated (in km) using
\begin{equation}
a_{1,2} = \frac{43200 K_{1,2} P \sqrt{1-e^2}}{\pi\sin i}.
\end{equation}
These values are given in Table~\ref{tab:phys_table} in units of AU. 
Combining these values of $a_1$ and $a_2$ with the angular semi-major 
axis of the relative orbit in Table~\ref{tab:sig_Sco_final}, the 
dynamical parallax was found to be $\pi_d = 5.76\pm0.68$\,mas. The 12\% 
uncertainty in the dynamical parallax is dominated by the contribution of 
the inclination -- the uncertainty in $\sin i$ is 11\%.

The distance to the system is therefore $174^{+23}_{-18}$\,pc, which is 
consistent within uncertainties with the {\em Hipparcos} value of 
$225^{+50}_{-35}$\,pc. An analysis of the {\em Hipparcos} data by
\citet{deBruijne99} produced a {\em secular} parallax of
$\pi_{\it sec} = 5.27\pm0.48$\,mas (or a distance of
$190^{+19}_{-16}$\,pc) with which the dynamical parallax (and consequent
distance) is in good agreement.

The members of nearby OB associations have been determined by 
\citet{deZeeuw99} from {\em Hipparcos} proper motion and parallax 
measurements. They concluded that $\sigma$~Sco is a secure member of the
Upper Scorpius (US) subgroup within the Sco OB2 association with a membership
probability of 94\%. Using  fig. 5 of \citealt{deZeeuw99}, the mean US 
parallax is 6.9\,mas with an estimated uncertainty of about 1.5\,mas 
(\citealt{deZeeuw99} give a `spread' value of 1.6\,mas). Therefore the 
dynamical parallax positions $\sigma$~Sco closer to the centre of 
US compared to {\em Hipparcos} and hence strengthens membership 
probability to the Upper Scorpius subgroup.

\begin{table*}
  \begin{minipage}{135mm}
  \centering
    \caption{Physical parameters of $\sigma$~Sco. The middle and bottom 
	     panels are the values for the primary and secondary respectively.}
    \label{tab:phys_table}
    \begin{tabular}{lcccc}
	\hline
Parameter     & Unit	  & This Work & Literature & Reference \\
	\hline
$a_1$         & AU   	  & $0.25 \pm 0.03$   &	-	    &  -	 \\	
$a_2$         & AU   	  & $0.38 \pm 0.07$   &	-	    &  -	 \\
$\pi$         & mas  	  & $5.76 \pm 0.68$   & $4.44\pm0.81$    & \citet{ESA97}  \\
	      & mas  	  & 		      & $5.27\pm0.48$    & \citet{deBruijne99}  \\
distance      & pc   	  & $174^{+23}_{-18}$ & $225^{+50}_{-35}$& \citet{ESA97}  \\
	      &		  &		      & $190^{+19}_{-16}$& \citet{deBruijne99} \\	
age           & Myr 	  &  10     	      &  5--14      & \citet{Brown98} \\
	      &		  &		      &    5	    & \citet{Preibisch02}  \\		
        \hline
${\cal M}$(B1~III) & M$_{\sun}$ & $18.4\pm5.4$     & -                &  -\\        		
$M_{V}$(B1~III)    & mag        & $-4.12\pm0.34$   &  -         &  -\\
$L$(B1~III)          & L$_{\sun}$ & $(2.9\pm0.8)\times 10^4$ 	      & - &  -\\
$R$(B1~III)          & R$_{\sun}$ & $12.7\pm1.8$     & -  &  -\\
	\hline	
${\cal M}$(B1~V) & M$_{\sun}$ & $11.9\pm3.1$     &  -         & - \\
$M_{V}$(B1~V)    & mag        & $-3.32\pm0.34$   &  -         & - \\
$L$(B1~V)  	      & L$_{\sun}$ & $(1.6\pm0.4)\times 10^4$ &  -         &  -\\
	\hline
    \end{tabular}
    \end{minipage}    
\end{table*}

%---------------------------------------------------------------------------%
\subsection{Component Masses}
\label{sig_masses}

The mass of the primary and secondary can be extracted from the orbital 
solution by combining Kepler's third law,   
\begin{equation}
\label{kepler_mass}
{\cal M}_1 + {\cal M}_2  = \frac{(a_1 + a_2)^3}{P^2},
\end{equation}
and the ratio of the component semi-major axes about the centre-of-gravity,
\begin{equation}
\label{mass_ratio}
\frac{{\cal M}_1}{{\cal M}_2}  = \frac{a_2}{a_1}.
\end{equation}
The calculated masses are in solar units when the semi-major axes are given 
in astronomical units and the period in years. Using the values determined 
in Sections~\ref{res} and \ref{sig_dis} the primary and secondary star
star masses were found to be $18.4\pm5.4$\,M$_{\sun}$ and 
$11.9\pm3.1$\,M$_{\sun}$ respectively. Once again the uncertainty in the
mass is dominated by the contribution of the inclination (propagated from 
the distance determination).

The expected mass of a B1~III star is $17.4$\,M$_{\sun}$ (interpolated from
values in \citealt{SKaler82}), while the analysis of HD92024 (a B1~III 
$\beta$~Cephei star with a nearly identical spectrum) 
by \citet{Freyhammer05} produced $15^{+3}_{-4}$\,M$_{\sun}$. Our 
determination of the primary star's mass in $\sigma$~Sco is in complete
agreement with these results. The mass of the primary is larger than the 
majority of members in the catalogue of Galactic $\beta$~Cephei stars 
\citep{Stankov05} where a mass of 12\,M$_{\sun}$ is the norm. The 
uncertainty in the mass is large, precluding the conclusion that 
the $\beta$~Cephei component in $\sigma$~Sco is one of the most massive 
examples of this type of pulsating star.

%---------------------------------------------------------------------------%
\subsection{$\beta$ Cephei Component Radius}
\label{sig_radius}

The analysis of the interferometric data yielded the uniform disc angular
diameter of the primary component. However, real stars
are limb-darkened and corrections are required to find the `true' angular
diameter from the uniform disc value. These corrections are discussed (and 
tabulated) in \citet{Davis00} and are dependent on the star's effective
temperature, surface gravity, chemical composition and the wavelength at 
which the uniform disc diameter was determined. Assuming solar chemical
composition, for $T_{\rmn{eff}} = 26\,150$\,K 
and $\log g = 3.85$ \citep{Linden88} at an observing wavelength of 700\,nm,
the limb-darkening correction factor given in \citet{Davis00} is
1.018 for the primary star. Using the dynamical parallax, the radius can
now be estimated to be $12.7\pm1.8$\,R$_{\sun}$, compatible with a B1~III 
star (interpolated) from \citet{SKaler82} and the analysis of the similar 
star HD92024 by \citet{Freyhammer05}. 

%---------------------------------------------------------------------------%
\subsection{Secondary Component}
\label{sig_sec}
Even though \citet{Mathias91} detected spectral features of the
secondary star, no estimation of its spectral type or luminosity class 
appears in the literature. 

Initial analysis of the interferometric data included the secondary angular 
diameter as a fitting parameter and a uniform disc radius of 
$11$\,R$_{\sun}$ was determined (but poorly constrained). Interpolating 
spectral type calibration tables in \citet{SKaler82}, stellar types B1~V 
\& B3~III had a mass that was consistent with that of the secondary 
(Section~\ref{sig_masses}). However, the fitted uniform disc radius was too 
large for a B1 dwarf star ($6.4$\,R$_{\sun}$) but was in accord for a B3 
giant ($10.8$\,R$_{\sun}$). On the other hand, single-star evolutionary 
tracks (Section~\ref{sig_lage}) predict vastly different ages for the two 
components and produce a more evolved state for a B3 giant secondary. 
Therefore a B1 dwarf classification was adopted and the interferometric 
data was reanalysed with the additional constraints on the likely radius
of the secondary taken into consideration. The results with this 
new fitting produced plausible values (absolute magnitudes and primary 
component radius more in line with literature calibrations, and similar ages) 
and without any significant change in the orbital parameters or quality of
the model-fit. 

Therefore, evolutionary tracks and the mass determined from the combination 
of interferometric and spectroscopic results imply a classification of
B1~V for the secondary. An approximate effective temperature of 
$T_{\rmn{eff}} = 25\,400 \pm2\,000 $\,K \citep{Lang91} is henceforth adopted 
for the secondary component.

%---------------------------------------------------------------------------%
\subsection{Component Magnitudes}

The absolute visual magnitude of the system can be determined from the 
distance and the apparent visual magnitude. The mean and standard deviation 
of extinction values found in the literature are $A_V = 1.23\pm0.20$\,mag 
(\citealt{Clayton93}; \citealt{Wegner02}; \citealt{Sartori03}; 
\citealt{deBruijne99}) and the the apparent visual magnitude of 
$\sigma$~Sco is $V= 2.88\pm0.02$\,mag \citep{Johnson66}. Combining 
these values with the distance from Section~\ref{sig_dis}, the absolute 
visual magnitude of the spectroscopic pair is 
$M_V({\rm B1~III + B1~V}) = -4.55$\,mag.

The absolute visual magnitudes of the components can now be determined from
the brightness ratio and $M_V({\rm B1~III + B1~V})$. However,  
the estimation of the brightness ratio in Section~\ref{res} was made
at a wavelength of 700\,nm and needs to be adjusted to the centre of the
$V$ band (550\,nm). Using blackbody radiation curves for 
$T_{\rmn{eff}} = 26\,150$\,K and $T_{\rmn{eff}} = 25\,400$\,K the adjustment 
factor to the 
measured brightness ratio was found to be 0.996 (i.e.\ a 0.004\,mag change 
to the magnitude difference) and hence, the $V$ band brightness ratio is 
$0.48\pm0.02$ (i.e.\ a 0.8\,mag visual magnitude difference). 
Hence, $M_V({\rm B1~III}) =-4.12 \pm 0.34$\,mag and 
$M_V({\rmn B1~V}) =-3.32 \pm 0.34$\,mag (the uncertainty estimates 
include the uncertainties in the distance, reddening and the
component brightness ratio). These values are in accord with the tabulations
of \citet{Panagia73} and \citet{Lang91}.

%---------------------------------------------------------------------------%
\subsection{Luminosity and Age}
\label{sig_lage}

Bolometric corrections found in the literature for a B1 giant are 
$-2.13$\,mag \citep{Panagia73} and $-2.43$\,mag \citep{Lang91}. Adopting the
mean value of $-2.28$\, mag, the luminosity of the primary component is
$(2.9 \pm 0.8)\times10^4$\,L$_{\sun}$. 
The secondary star has a proposed classification of a B1 dwarf which has 
bolometric corrections of $-2.23$\,mag  and $-2.70$\,mag from \citet{Panagia73}
and \citet{Lang91} respectively. The luminosity of the secondary is
$(1.6 \pm 0.4)\times10^4$\,L$_{\sun}$ using a mean bolometric correction
of $-2.47$\,mag. Both the primary and secondary luminosities are consistent
(within uncertainties) with the tabulations of \citet{Panagia73} and 
\citet{Lang91}.
\begin{figure}
    \includegraphics{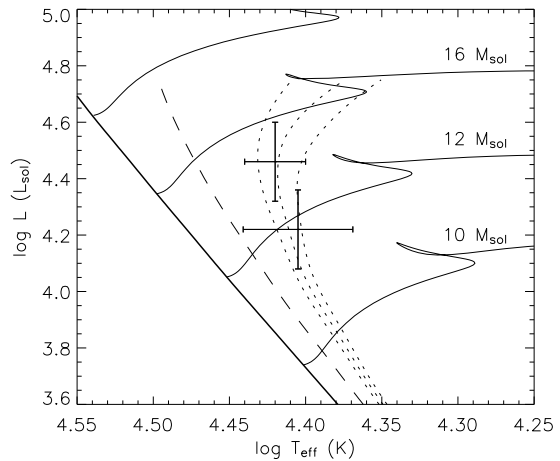}     
    \caption{Single-star evolutionary tracks of 
		 {\protect \citet{Claret04}} with calculated isochrones of 
		 9, 10, 11\,Myr given as dotted lines with the dashed line 
		 corresponding to the US group age of 5\,Myr {\protect
		 \citep{Preibisch02}}. The locations of the components are 
		 marked -- the lower cross being the secondary.}
        \label{fig:sig_Sco_evo}	
\end{figure}

The ages of the component stars can now be estimated using the single-star
evolutionary models of \citet{Claret04}. The positions of the stars
in the HR diagram (Fig.~\ref{fig:sig_Sco_evo}) are marked with crosses, 
the lower-luminosity one being the secondary star (with larger error
bars reflecting the considerable uncertainty in its effective temperature). 
Isochrones of 9, 10 and 11 M\,yrs have been calculated and shown as dotted 
lines with the dashed line corresponding to an age of 5\,Myr. The 
evolutionary-model masses of both stars\footnote{The initial erroneous 
classification of the secondary as a B3 giant (Section~\ref{sig_sec}) 
produced a location in the bottom right-hand corner of 
Fig.~\ref{fig:sig_Sco_evo} corresponding to an evolutionary-model mass of 
approximately 8\,M$_{\sun}$.} are compatible with the determination in 
Section~\ref{sig_masses}, even though the primary has a lower 
evolutionary-model mass than the measured value.

The age of the system is estimated to be 10\,Myr and from the isochrones
coeval formation of the two components is confirmed. The age range of the 
entire Sco OB2 association (US + Upper Centaurus-Lupus + Lower 
Centaurus-Crux) is 5--14\,Myr \citep{Brown98}, consistent with the age 
determined for $\sigma$~Sco. However, a recent exploration of the stellar 
population of US by \citet{Preibisch02} suggests an age of 5\,Myr for the 
group. They propose that the US members are the result of a shock wave 
passing through a molecular cloud 5--6\,Myrs ago causing a burst in star 
formation and the shock wave being from a supernova in the Upper 
Centaurus-Lupus group about 12\,Myr ago.
However, $\sigma$~Sco is an `outlier' in the colour-magnitude diagram
of \citet{Preibisch02} (and \citealt{deBruijne99}) who explain the 
deviation to be a result of its binary and pulsation characteristics.
Hence there is an implication that $\sigma$~Sco may have formed prior to 
the other members of the US group but the effect of binarity and 
observational uncertainties must be investigated further before the 
evolutionary status of $\sigma$~Sco can be confirmed.

%---------------------------------------------------------------------------%
%
%		        	SUMMARY
%
%---------------------------------------------------------------------------%

\section{Summary}
\label{sig_summ}

The first complete orbital solution for $\sigma$~Sco, based on 
interferometric measurements with SUSI, is presented. 
In combination with the only double-lined radial velocity measurements, the 
dynamical parallax and distance to $\sigma$~Sco have been determined
and shown to be consistent with previous estimates. Furthermore, the
masses of the components have been determined which, in combination with
evolutionary tracks, allows the first proposed classification of the
secondary as B1~V.

Using calibrations found in the literature, the absolute visual magnitude, 
luminosities and mass of the component stars are found to be consistent with 
types B1~III and B1~V for the primary and secondary respectively. 
Furthermore, the radius of the primary, determined from the interferometric 
angular diameter and distance, is in accord with a B1 giant classification.

The 10\,Myr age of the spectroscopic pair, estimated from the single-star
evolutionary models of \citet{Claret04}, is in agreement with previous
estimates of the Sco OB2 
association as a whole (5--14\,Myr) but not with other members of the Upper 
Scorpius group (5\,Myr). Further information (in the form of improved 
component characteristics) is needed to either resolve this discrepancy or 
show that $\sigma$~Sco formed earlier than the remainder of the Upper 
Scorpius group.

%---------------------------------------------------------------------------%
%
%			    ACKNOWLEDGEMENTS
%
%---------------------------------------------------------------------------%

\section*{Acknowledgments}

This research has been jointly funded by The University of Sydney and the 
Australian Research Council as part of the Sydney University Stellar
Interferometer (SUSI) project. We wish to thank Brendon Brewer for his 
assistance with the theory and practicalities of Markov chain Monte Carlo 
Simulations. Andrew Jacob and Stephen Owens provided assistance during 
observations. The SUSI data reduction pipeline was developed by Michael 
Ireland. JRN acknowledges the support provided by a University of Sydney 
Postgraduate Award. This research has made use of the SIMBAD database,
operated at CDS, Strasbourg, France.

%---------------------------------------------------------------------------%
%
%			    	REFERENCES
%
%---------------------------------------------------------------------------%

\bsp

\label{lastpage}

\end{document}